\documentclass[12pt]{article}

\newif\ifSubmit
\Submittrue

\usepackage{cite}
\usepackage{graphicx}
\usepackage{times}
\usepackage{color}
\usepackage{comment}
\usepackage{url}

\ifSubmit
\fi

\title{\textbf{Technical Article:} \\ On the Optical Accuracy of the \textit{Salvator Mundi}}

\author{
Marco (Zhanhang) Liang (computer scientist, artist, researcher), \\
University of California, Irvine, Dept. of Computer Science, \\
Irvine, CA 92697-3435, USA.
Email: \texttt{$<$zhanhanl@uci.edu$>$}.
\\[5pt]
Michael T.~Goodrich (computer scientist, researcher), \\
University of California, Irvine, Dept. of Computer Science, \\
Irvine, CA 92697-3435, USA.
Email: \texttt{$<$goodrich@uci.edu$>$}. \\
ORCID:0000-0002-8943-191X.
\\[5pt]
Shuang Zhao (computer scientist, researcher), \\
University of California, Irvine, Dept. of Computer Science, \\
Irvine, CA 92697-3435, USA. 
Email: \texttt{$<$shz@ics.uci.edu$>$}. \\
ORCID:0000-0003-4759-0514.
}

\date{}

\newcommand{\marco}[1]{{\color{blue} [MARCO: #1]}}

\newcommand{\Leonardo}{{\it Leonardo?}}

\begin{document} 


\baselineskip20pt


\maketitle 

\begin{abstract}
A debate in the scientific literature has arisen regarding whether the orb depicted in \textit{Salvator Mundi}, which has been attributed
by some experts to Leonardo da~Vinci,
was rendered in a optically faithful manner or not.
Some hypothesize that it was solid crystal while others hypothesize that it was hollow, with competing explanations for 
its apparent lack of background distortion
and its three white spots.
In this paper, we study the optical accuracy of the \textit{Salvator Mundi} using physically based rendering,  
a sophisticated computer graphics tool that produces optically accurate 
images by simulating light transport in virtual scenes.
We created a virtual model of the composition centered 
on the translucent orb in the subject's hand.
By synthesizing images under configurations that vary illuminations and orb material properties, we tested whether it is optically possible to produce an image that renders the orb similarly to how it appears in the painting.
Our experiments show that an optically accurate rendering qualitatively matching that of the painting is indeed possible using materials, light sources, and scientific knowledge available to Leonardo da~Vinci circa 1500.
We additionally tested alternative theories regarding the composition of the orb, such as that it was a solid calcite ball, which provide empirical evidence that such alternatives are unlikely to produce images similar to the painting,
and that the orb is instead hollow.
\end{abstract}


\textit{Salvator Mundi} is a painting of Christ (see Fig.~\ref{fig:salvator_mundi}A) dated to ca.~1500 that has been attributed
by some experts to Leonardo da~Vinci.%
\footnote{
	Although many believe that this painting was created by Leonardo da~Vinci, the authorship is still being debated.
	In this paper, we use ``\Leonardo{}'' to reflect this uncertainty.
}
Although centuries old, it was rediscovered, restored, 
and authenticated only recently.
When it sold in 2017 for \$450.3 million, 
it became the most expensive painting ever sold\cite{artnews17}.

Since its rediscovery, a debate has arisen as to whether the
translucent orb in the painting was rendered accurately.
Even though the painting dates to a period when Leonardo da~Vinci 
was studying optics, 
several observers have commented that the orb is not
rendered as it should were it a solid glass or crystal orb,
which would invert and magnify the subject's robe behind the orb and obscure the subject's palm.
For example, Isaacson\cite{isaacson} hypothesizes in his biography,
\textit{Leonardo da Vinci}, that Leonardo deliberately
rendered the orb inaccurately, while
Kemp\cite{kemp1}, who helped authenticate the \textit{Salvator Mundi},
writes in \textit{Nature} 
that the orb ``glistens with points of light'' that are not ``spherical
bubbles found in glass, but are the kind of cavity inclusions (small
gaps) that appear in some specimens of rock crystal and calcite.''
Kemp further writes that Leonardo ``observed the double refraction
produced by calcite.'' 
In a follow-up \textit{Nature} correspondence,
Noest\cite{noest} questions Kemp's interpretation, noting a lack of
optical distortion in the orb, and he hypothesizes that the specks on the
orb's surface were instead painted on the orb.
Kemp\cite{noest} replies, ``Leonardo did not aspire to
represent his subjects as if he were a scientist recording natural
phenomena.''
Such correspondences have not
settled the issue, however, and this debate has 
continued\cite{alberge,kinsella,reporters}.

Fortunately, we can use scientific tools to address 
such controversies\cite{Abraham2010,Arecchi2006,criminisi2005bringing,Hockney:00,Stork,Stork2,doi:10.1001/jamaophthalmol.2018.3833}.
For example, the technique of 
\emph{inverse rendering} uses tools from computer graphics and applied optics to infer
scene information from 
photographs\cite{Boivin:2002:2158-6330:268,Marschner:1997:2166-9635:262,Ramamoorthi:2001:SFI:383259.383271,Yu:1999:IGI:311535.311559}.
Here, although it is not a photograph,
we nevertheless apply a type of inverse rendering to
the {\it Salvator Mundi} using 
a physically based renderer (PBR), which is a
sophisticated Computer Graphics tool designed to 
simulate the physics of light flow through a three-dimensional virtual
scene to produce an optically accurate image\cite{pbr}.
By comparing synthetic images produced by a PBR with the painting,
we qualitatively tested various hypotheses regarding the optical
accuracy of the painting with respect to various materials and light
sources that the artist might have used.
For instance, 
we used the Mitsuba PBR\cite{mitsuba} 
to produced renderings that are faithful to the 
painting (e.g., see Fig.~\ref{fig:salvator_mundi}B), 
which provide evidence regarding possible materials
and lighting conditions used by \Leonardo{}.

\section*{Experimental Setup}

We built a virtual scene with an approximated geometry gleaned from the painting and experimented with a variety of configurations for the orb.
Our experiments explored possible hypotheses regarding material properties of the orb, with particular interest in possible explanations for the lack of optical distortion by the orb.

%
\section*{Virtual Scene Setup}
Since our main goal was to test hypotheses regarding the material property of the orb, we depict of the scene geometry using a rough approximation for the subject's body along with more detailed representations for the orb and the hand holding it.
Specifically, we sculpted a three-dimensional relief as a proxy geometry for the subject, and used the painting (with the orb removed) as a color texture for this relief.
Then, we placed a three-dimensional orb as well as the hand holding this orb in front of the relief model, 
as shown in Fig.~\ref{fig:setup}.
We used the subject's left hand as a reference and scaled the orb and relief accordingly to match the relative size and position to the painting.
%
%
This results in the orb with 6.8 cm in radius located 25 cm away in front of the subject.

Additionally, we fine-tuned our model to further improve its visual quality as follows.
We refined the geometry of the orb-holding hand using Maya, a 3D modeling and animation software, to make it touching the orb softly while avoiding any overlap between the two models.
We also slightly adjusted the light transmission rate of the orb material 
(as a dielectric) to resemble the background-darkening effect from the painting.
Lastly, to better reproduce the overall smooth appearance, we applied Gamma correction to both the relief texture and the final rendered image.

Besides geometry, illumination and viewing configurations are also key to visual appearance.
Using visual cues including the brightness gradients on the subject's face, chest, and hands as well as the shadows under the chin, we conclude
that the painted scene to involve a strong directional light source from the above and introduced a similar light  in our virtual scene.
Specifically, we used a few directions around a main one (instead of using a single directional light) to slightly soften the shadow boundaries.
To reproduce the ambient illumination (which reduces the contrast between well-lit and shadowed areas), we used a dim constant environmental light.
From a photographic point of view, the directional light acts as the ``main light'' of the scene while the dim environmental light as the ``fill light''.
Lastly, we used a perspective viewpoint 
(which is anachronistically called a ``camera'' in the context of PBR) 
that is located 90 cm away and points straight at the subject to simulate the typical viewing configuration in a studio.

\section*{Material Properties of the Orb}
\subsection*{Solid or Hollow?}
With the virtual scene ready, we tested whether the orb was
solid by comparing renderings of a solid and a hollow orb.
As noted by others\cite{isaacson,noest},
a solid orb bends light as 
a convex lens would, which would invert and magnify the image 
of the robe behind the orb (e.g., see~Fig.~\ref{fig:solidity}A).
This effect persists regardless of the orb's material 
(as long as it is optically denser than air).
A hollow orb, in contrast, does not cause such distortion. 
As can be seen in Fig.~\ref{fig:solidity}B,
the folds of the robe are not distorted or inverted
by a hollow orb.
Based on this comparison, we conclude that the orb was most likely hollow 
(assuming the painting follows the physics of light transport).

Knowing that the orb was likely hollow, we experimented 
on how such an orb alters the shape of lines on the boundary.
We found that when a hollow glass ball is placed 
directly in front of a straight line, relative
to the viewing ``eye,'' the optics imply that we can see that
the straight line appear to pass into the ball without distortion 
when the center of the ball is on this line, but this lack of distortion
is not true for lines that do not pass through the ball's center.
For example, suppose we were to place
a hollow glass ball directly above three 
parallel straight lines, with a camera viewing from top,
As illustrated in Fig.~\ref{fig:experiment}A.
The line in the middle, which passes through the center of the 
ball from camera's viewpoint, 
preserves its continuity on the boundary, 
while the other two have curved distortions
(shown in Fig.~\ref{fig:experiment}B).
The alignment of three things---the straight line, viewing ``eye'', and center of the hollow ball---makes it possible that a line passes through a hollow ball without interruption. 
This is true even for solid balls (with arbitrary thicknesses), as illustrated in Fig.~\ref{fig:experiment}C.
Another key factor in having a connected line as it is refracted
and reflected with respect to a hollow ball
is to keep the line straight outside the ball. 
If we bend the middle line to the right outside the ball, 
for example, 
it becomes disconnected on the boundary(Fig.~\ref{fig:experiment}D).

To test whether \Leonardo{} may have utilized these optical 
devices to his composition and placement of the connected folds, 
we analyzed the painting, especially the composition and setup 
of the folds in the subject's robe (Fig.~\ref{fig:holloworb}A).
In the painting, the drape of the subject's clothes introduces
five prominent folds that pass through the orb's upper-right boundary, 
among which the leftmost one has a cast shadow wider than the others. 
By drawing straight lines along the edge of these folds (as indicated with solid white line in Fig.~\ref{fig:holloworb}B), 
we found that, except for the leftmost fold, 
the four other folds converge to a point, which turns out to be the center of the orb.
This convergent point, or center of the orb, is a hollow dot with light grey color and smaller 
than three white spots. 
Moreover, the folds keep straight for a distance outside the orb before 
they curve towards subject's left shoulder. 
Based on this 
realization of this geometry 
of the folds in the painting as essential in having 
connected folds on orb's boundary, we refined the 3-dimensional relief 
in our model to 
assure the folds were straight and radiative.
Furthermore, this analysis suggest that \Leonardo{} understood these
optical properties of hollow balls and how to avoid distracting optical 
distortions from the rendering of the folds of the subject's robe.

\subsection*{Thickness}
With the above findings, we further explored the thickness of the orb 
in the painting 
by comparing the smoothness of the folds going across the orb's boundary. 
Given the importance of the position, we aligned the viewing 
``eye'' (camera), the orb's center, and convergent point of the folds.
As the hollow orb was likely made via glass blowing with the help of a hemisphere mold, we modeled it as hollow with a refractive index of 1.51714. 
With these settings, we determined a thickness relating to the size and position of orb.
In particular, for a hollow glass orb that is 6.8 cm in radius, 
and positioned so that the folds of the robe follow lines approaching 
the center of the orb (relative to the viewpoint), 
we determined that the maximum thickness of the orb could be 1.3 mm 
without producing noticeable abruptions on the folds that traverse the boundary (Fig.~\ref{fig:holloworb}C). 

In the rendering, the leftmost fold, whose extended edge fails to pass through the 
convergent point, is distorted by the boundary, while the four other folds cross 
the boundary smoothly (Fig.~\ref{fig:holloworb}D).
\Leonardo{} chose to blur this distorted edge in the painting, but rendered the
others accurately, suggesting that the artist was aware of the optics of how
straight lines appear when, relative to the viewing ``eye,''
they cross the boundary of a hollow glass sphere.
Interestingly, the cast shadow of the leftmost fold has its outer region, 
which is further-away and soft-edge, converges to the center, 
and has a smooth transition across the orb's boundary (as denoted by dotted line in Fig.~\ref{fig:holloworb}B).

To further illustrate how the thickness and placement 
of the orb relative to the folds of the subject's robe
impact how the folds of the robe appear, 
we altered the orb's thickness and the orb's position to show
resulting distortions. 
We first rendered a hollow orb with 2.6 mm in thickness, 
and found that the folds are intercepted by a dark layer, which is a reflection of the ambient environment (i.e., the room in which the subject sits), as shown in Fig.~\ref{fig:alternative}A.
As for orb position, we horizontally moved the hollow orb 1 cm to the left relative to the viewing ``eye,'' and rendered the result, which shows that in this case the folds appear disconnected
(Fig.~\ref{fig:alternative}B).

%
%
\section*{Alternative Theories}
In addition to exploring practical orb thickness, 
we used physics-based simulation to test an alternative theory 
regarding the orb.
This theory, which was proposed by Martin Kemp, argues that 
the orb is made of solid calcite and that the two distinct contours 
on the heel of the subject's hand are due to birefringence~\cite{kemp1}. 
We rendered the birefringence of such an orb by averaging two solid orb images 
rendered with an extraordinary index of refraction of 1.486 
and an ordinary index of refraction of 1.658.
Our physics-based simulation (Fig.~\ref{fig:birefringence}) shows that 
this is unlikely, since the solidity of the orb would cause visually significant 
distortion (of the robe and the palm behind the orb) that is more prominent 
than the double-contours effect and these effects
are clearly absent from the original painting.

\section*{Discussion}
Given our experimental results, a natural question is whether 
Leonardo da~Vinci had access in 1500 to the materials,
light sources, and scientific knowledge of optics represented in configurations
we used with the Mitsuba PBR to generate images
similar to the painting, \textit{Salvator Mundi}.
Fortunately, besides being an artist,
Leonardo was also a scientist who kept copious notes, much of
which have survived, and these notes shed light on this
question.
Indeed, a recent volume,
{\it Leonardo da Vinci and Optics},
is devoted to studying the relationship between optics 
and the paintings of Leonardo~\cite{optics13}.

Furthermore,
in 1883, Richter\cite{richter} published a compilation
of Leonardo's notes, including drawings and English translations.
These notes show, for example,
that Leonardo had an understanding of 
light refraction (e.g., no.~75, Fig.~\ref{fig:notes}A),
glass and crystal materials and
diffused, direct, and reflected light (e.g., no.~118, Fig.~\ref{fig:notes}B),
the relative position of reflections on a round body
(e.g., no.~134, Fig.~\ref{fig:notes}C),
how light can be directed through a ``window''
(e.g., no.~146, Fig.~\ref{fig:notes}D),
reflected colors (e.g., no.~283, Fig.~\ref{fig:notes}E),
how to create a semi-diffused light source using paper and a candle
(e.g., no.~524, Fig.~\ref{fig:notes}F),
and even how light reflects from a concave mirror (e.g., Fig.~\ref{fig:notes}G).
Thus, these notes provide
evidence that Leonardo had access to the materials, light sources,
and scientific knowledge necessary to create configurations that our
experiments show can produce images similar to how the orb
is rendered in \textit{Salvator Mundi}.

%



\section*{Acknowledgments}
SuperDasil provided the hand model on BLEND SWAP.
This research was supported in part by
NSF-BSF grant 1815073.
MZL designed the geometric models and performed the experiments.
SZ provided supervision, insights regarding conical light
paths, and PBR expertise.
MTG provided supervision,
originated the research question, and guided research progress.
All three authors contributed to the writing of the paper.
Data for our configurations and simulations can be found at \url{https://github.com/salvator-mundi/three-white-spots}.

%


\clearpage

%
%
\ifSubmit
	\begin{figure}[t]
	\centering
	\includegraphics[width=\textwidth, trim = 0in 1.8in 0in 0in, clip]{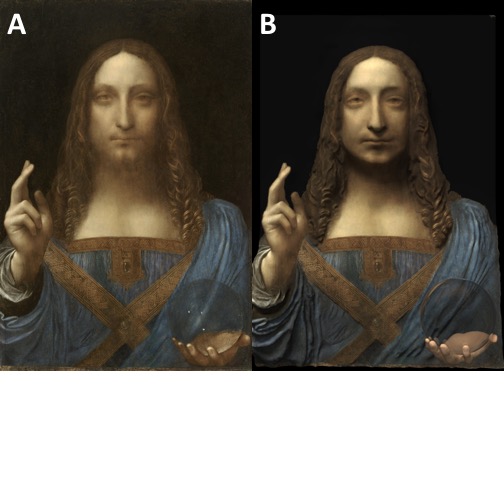}
	\caption{
	(A) The \textit{Salvator Mundi} (public domain image). 
	(B) A PBR rendering of the
        \textit{Salvator Mundi} using a hollow orb.
        }
	\label{fig:salvator_mundi}
\end{figure}
\fi

%
%
\ifSubmit
	\begin{figure}[t]
	\centering
	\includegraphics[width=\textwidth, trim = 0in 3.5in 0in 0in, clip]{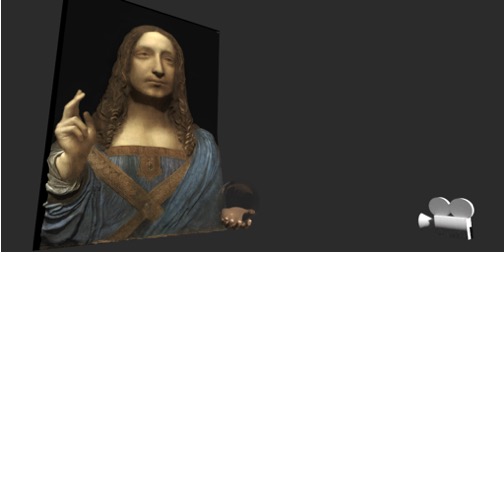}
	\caption{
	Virtual scene setup where an orb-holding hand model is positioned in front of the subject's relief, which is textured with a modified version of the painting. 
	}
	\label{fig:setup}
\end{figure}
\fi

%
%
\ifSubmit
	\begin{figure}[t]
	\centering
	\includegraphics[width=\textwidth, trim = 0in 3.5in 0in 0in, clip]{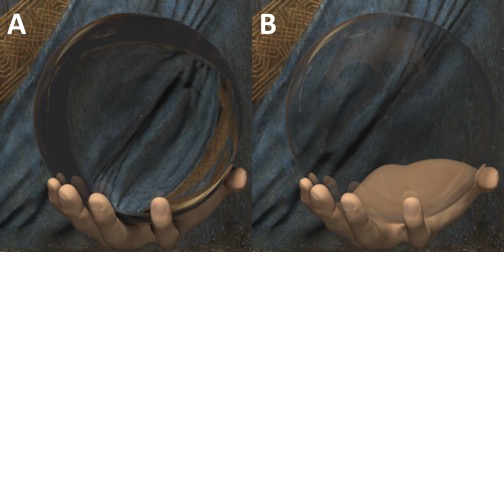}
	\caption{
	(A) Rendering of a solid orb. 
	(B) Rendering of a hollow orb.
	}
	\label{fig:solidity}
\end{figure}
\fi

%
%
\ifSubmit
	\begin{figure}[t]
	\centering
	\includegraphics[width=\textwidth]{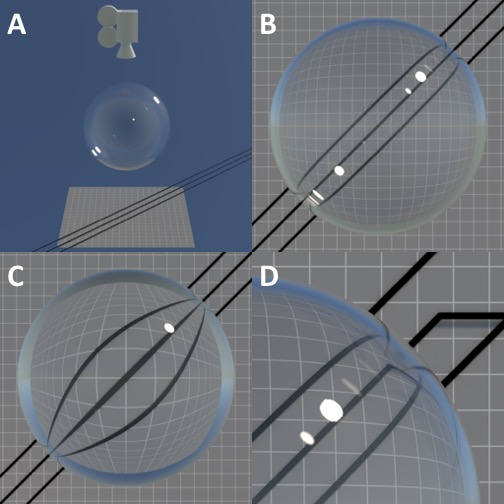}
	\caption{
	(A) Scene setup for the experiment, where a glass ball is above three lines and a camera views from the top. 
	(B) A hollow glass ball in front of three straight lines. 
	(C) A solid glass ball in front of three straight lines. 
	(D) A hollow glass ball in front of three lines. Outside the ball, the line in the middle bends to the right .
	}
	\label{fig:experiment}
\end{figure}
\fi

%
%
\ifSubmit
	\begin{figure}[t]
	\centering
	\includegraphics[width=\textwidth]{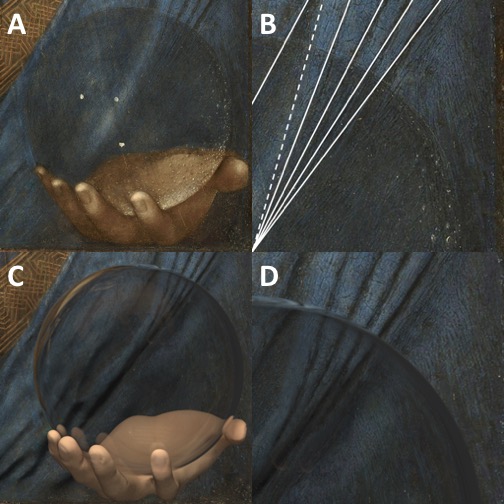}
	\caption{
	(A) The orb from the \textit{Salvator Mundi} for reference. 
	(B) Illustration of the folds on orb's boundary, with white solid lines drawn along folds' edge, 
	and a cast shadow denoted by a dotted line.
	(C) Rendering of a hollow orb with 1.3 mm in thickness.
	(D) Upper-right boundary of the hollow orb from (C).	
	}
	\label{fig:holloworb}
\end{figure}
\fi

%
%
\ifSubmit
	\begin{figure}[t]
	\centering
	\includegraphics[width=\textwidth, trim = 0in 3.5in 0in 0in, clip]{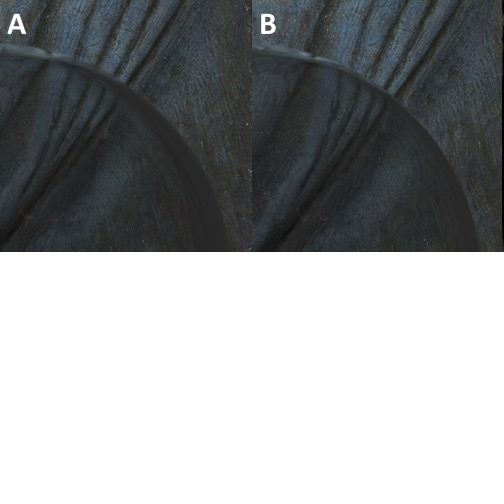}
	\caption{
	(A) A hollow orb with 2.7 mm in thickness. 
	(B) A hollow orb with 1 cm shift to the left, relative to the viewing ``eye''. 
	}
	\label{fig:alternative}
\end{figure}
\fi

%
%
\ifSubmit
	\begin{figure}[t]
	\centering
	\includegraphics[width=.5\textwidth, trim = 0in 3.5in 3.5in 0in, clip]{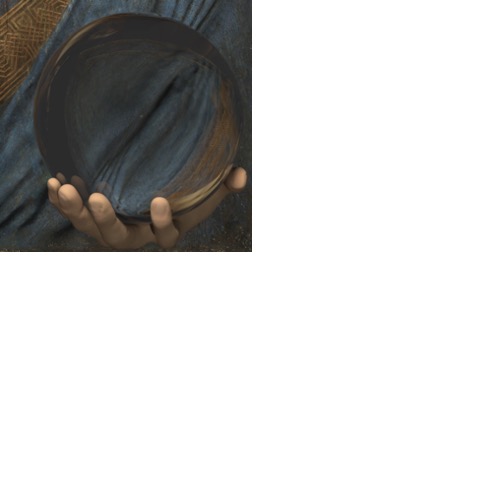}
	\caption{
	Rendering of a solid calcite orb with birefringence.
	}
	\label{fig:birefringence}
\end{figure}
\fi

%
%
\ifSubmit
\begin{figure}[t]
	\centering
	\includegraphics[width=\textwidth, trim = 0.5in 0in 0.2in 0.1in, clip]{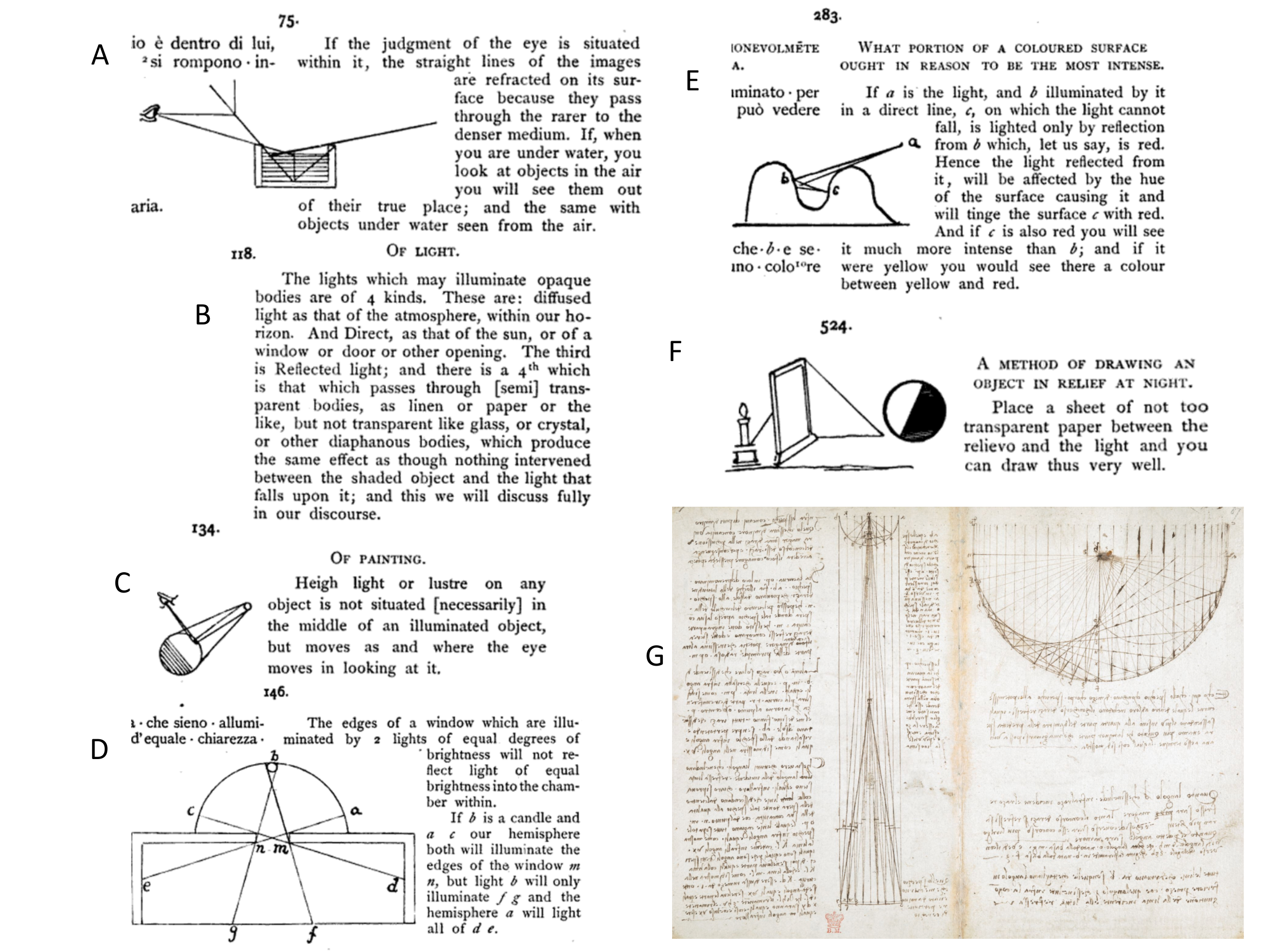}
	\caption{(A-F) Excerpts from the notes of Leonardo da~Vinci,
	as compiled and translated by Richter~\cite{richter}. (G) 
	Notebook of Leonardo da~Vinci (The Codex Arundel): 
	Studies of reflections from concave mirrors.
	British Library Arundel MS 263, f.87r.}
	\label{fig:notes}
\end{figure}
\fi

\end{document}